\documentclass[fleqn,twoside]{article}
\usepackage{espcrc2}

\usepackage{graphicx}
\usepackage[figuresright]{rotating}

\newcommand{\AmS}{{\protect\the\textfont2
  A\kern-.1667em\lower.5ex\hbox{M}\kern-.125emS}}
  \newcommand*{\meg}            {\mu \to e \gamma}
\newcommand*{\megsign}        {\mu^+ \to e^+ \gamma}

\newcommand*{\egamma}         {E_{\gamma}}
\newcommand*{\epositron}      {E_{\rm e}}
\newcommand*{\tegamma}        {t_{{\rm e}\gamma}}

\newcommand*{\dtheta}         {\theta_{e \gamma}}
\newcommand*{\dphi}           {\phi_{e \gamma}}
\newcommand*{\nsig}           {N_{\rm sig}}
\newcommand*{\nenunu}         {N_{e \nu \bar \nu}}
\newcommand*{\nrd}            {N_{\rm RMD}}
\newcommand*{\nbg}            {N_{\rm BG}}
\newcommand*{\nobs}           {N_{\rm obs}}
\newcommand*{\nubar}           { \bar \nu }

\title{ Recent MEG results}

\author{G.Cavoto\address[INFN]{Istituto Nazionale di Fisica Nucleare, 
        Sezione di Roma, \\ 
        Piazzale A.Moro,2   00185 Roma, Italy}%
                }
       
\begin{document}

\begin{abstract}
  New results of a search for the ultra-rare decay $\mu \to e \gamma$ by the MEG collaboration are reported 
  in this contribution. The data were taken during  2009  and correspond to 
   approximately 6.5 10$^{13}$ muon stopped on target. A maximum likelihood analysis  sets an  upper limit at 90\% C.L.  
    on the branching ratio, BF($\mu \to e \gamma$) $< $1.5 10$^{-11}$.  The results presented here are  preliminary.
\vspace{1pc}
\end{abstract}
\maketitle

\section{INTRODUCTION}
Lepton Flavour Violation (LFV)  processes  in the charged sector
 are highly suppressed  in the Standard Model (SM). This is due to 
 the non-zero   neutrinos  masses  and  to their mixing \cite{Strumia:2006db}. 
 A SM prediction for  BF($\mu \to e \gamma$) is approximately 10$^{-54}$\cite{Petkov,Cheng:1980tp}.
  Such low BF is  virtually unobservable and therefore any positive signal of such reaction
 would be a clear evidence of New Physics.  Supersymmetric models  - on the other hand - can generate 
 flavour mixing effects \cite{Barbieri:1994pv,Barbieri:1995rs,Hisano:1996qq,Calibbi:2006nq,Calibbi:2009pv}  at a level 
 they can be  investigate  by experiment of the current generation.
   
  We report here on the results of a search for the LFV  decay $\mu^+ \to e^+ \gamma$, 
  based on data collected
during the 43 days  of data acquisition in November and
December 2009 by the MEG experiment. The MEG experiment is operated  at the 590$\,$MeV proton ring cyclotron facility of the
Paul Scherrer Institut (PSI), in Switzerland. The muons  originated from a target  stop in the experiment with  rate  adjusted  to be 2.8 $10^7$ Hz.
 In  total  approximately 6.5 10$^{13}$ muons were  stopped on the  target  and 22 M triggered events  acquired.

   The smallest  limit for the branching ratio BR$(\mu \to e \gamma) \leq 1.2
\times 10^{-11}$ (90$\%$~C.L.), was  set by the MEGA
experiment~\cite{MEGA}. The MEG collaboration recently published a limit of the same order of magnitude
 based on the analysis of the first data collected during 2008\cite{Adam:2009ci}. 
 
 \section{THE MEG EXPERIMENT}

  The $\mu^+ \to e^+ \gamma$  process is characterized by a simple two-body final
state, with the positron and photon being emitted  in time and
 back-to-back in the rest frame of the muon, each with an
energy equal to half the muon mass.
 There are two major sources of background,  the dominant being the 
accidental coincidences between a high energy positron from
the  principal  decay $\mu^+ \to e^+ \nu_e \bar\nu_\mu$
(Michel decay)  and a high energy photon from positron annihilation-in-flight or bremsstrahlung
 or   from the  radiative muon decay (RMD) $\mu^+ \to e^+ \nu_e \bar\nu_\mu\gamma$.
Moreover,   a small fraction of RMD  is a source of correlated background, producing a  high energy photon  in time with a positron.

 The MEG experiment  layout  combines the use of a  continuous muon beam and  a high precision detector with excellent spatial, temporal and energy
resolutions. The MEG detector covers 10\% of the total solid angle and comprises a photon detector and a positron spectrometer.
A schematic of the experiment is shown in Fig.1.

 \begin{figure}[h]
 \label{fig:layout}
\includegraphics[scale=0.25]{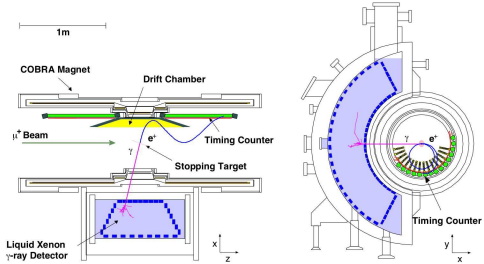}
\caption{Schematic drawing of the MEG detector. $z$ axis is chosen along the muon beam direction.}
\end{figure}

A beam of surface muons of 28$\,$MeV/c   is   separated by  the eight times higher positron contamination
 with a Wien filter and then stopped in a thin, slanted
 polyethylene  target, placed at the centre of
the detector.  The use of a helium environment   ensures  minimal
multiple scattering for both the muons and the out-going positrons and limits  
photon background production in the acceptance region.

Positrons originating from muon decay are analyzed in the COBRA
(COnstant-Bending-RAdius) spectrometer consisting of a thin-walled
superconducting magnet with a gradient magnetic field  and a tracking
system of 16 radially aligned modules of low-mass ($2.0\times10^{-3}$\,$X_0$ in total) drift chambers (DCH).
 The gradient magnetic field  (1.27-0.49 T)
has larger acceptance for  higher momentum particles and minimizes 
the track-length of charged particles compared to a uniform field.
 
 Two fast scintillator bars  arrays  (TC) are  placed at each end of the spectrometer and  
are  read  at either end by a fine-mesh photomultiplier tube (PMT),  providing  positron timing information
 and its  impact point position. 
 
   Photons are detected in a 900~litre homogeneous volume of liquid xenon
(LXe)  by measuring  the  emitted scintillation light  with 846 PMTs. LXe has a  fast response,
large light yield and short radiation length and allows to measure the  total energy released by the
$\gamma$-ray as well as the position and time of its first
interaction.
 
  To select matched photon and positron candidates in a high rate,
continuous beam environment and store sufficient information for
offline analysis requires a well matched system of front-end
electronics, trigger processors and data acquisition (DAQ) software. 
The front-end electronics signals (2748) are actively split and go
to both the trigger and the in-house designed waveform digitizer
boards,  based on the multi-GHz domino ring sampler chip (DRS). 
This  system achieves
an excellent pile-up recognition, together with  superior timing and
amplitude resolutions, compared to conventional schemes.
The trigger is based on fast information from both TC and LXe  requiring an energy deposit in the photon
detector in an interval around $52.8\,$MeV, a time coincident
positron hit on TC  within 20 ns and a rough collinearity of the
two particles. 
The typical signal event rate was $5$~Hz, and the total
DAQ rate was $6.5$~Hz, with an average livetime of $84\%$. The trigger efficiency on signal events is $84$\%.

\section{CALIBRATIONS and  RESOLUTIONS}

\begin{figure}[h]
\includegraphics[scale=0.4]{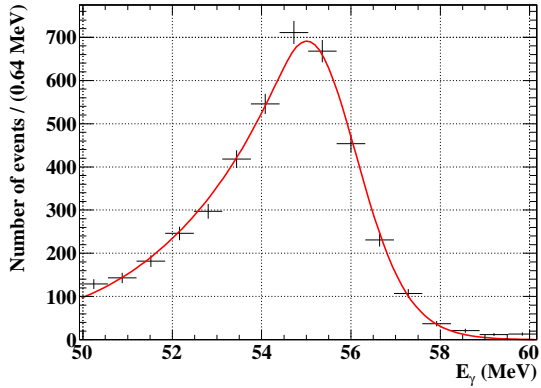} 
\caption{Reconstructed  $55$ MeV photon energy spectrum from   $\pi^- p \to \pi^0(\gamma\gamma) n$ CEX process. 
The shown fit  function contains a Gaussian component to fit the right-hand side part.
  This spectrum includes only photons with a conversion point distant more than  2 cm  from the LXe front-face.}
\label{fig:CEXreso}
\end{figure}

  Calibration and monitoring of the apparatus are key ingredients to reach the sensitivity of $10^{-12}-10^{-13}$.
  In particular the LXe PMT were constantly calibrated with LEDs and $\alpha$-sources. The energy scale and resolution 
 was  evaluated with differences processes in a range from few MeV to hundred MeV using  (p,$\gamma$) reaction excited with a 
  dedicated proton Cockroft-Walton accelerator and a  $\pi^- p$ charge exchange and radiative capture reaction (CEX).
     The relative time between the LXe and TC was monitored using (p,2$\gamma$) events. 
   The light yield and the energy scale was found to be stable well below 1\% level.
  In Fig.\ref{fig:CEXreso} the  energy spectrum for   $55$ MeV photons from CEX process is shown, demonstrating that  a resolution $\frac{\sigma_E}{E}$ = 2.1 \% 
  has been achieved.

  After detailed calibration procedures  DCH and TC    show good intrinsic position and time resolution for the positron. Measurements of radial positions of a
  positron track  have resolution of 250 $\mu$m while $z$ resolution can be described by a double Gaussian function with $\sigma$ =  600 $\mu$m for 70\% of the tracks.
  TC  positron time resolution averaged over the various bars is about 70 $ps$.

\section{DATA ANALYSIS and RESULTS}

 Events falling into a pre-defined window
(blinding-box), containing the signal region on the $\gamma$-ray
energy and the time difference between the $\gamma$-ray and the
positron, are  written to a separate
data-stream, in order to prevent any bias in the analysis procedure.
Only the events outside the blinding-box are used for optimizing the
analysis parameters and for studying the background.

A candidate $\mu^+ \to e^+ \gamma$ event is characterized by the
measurement of five kinematic parameters: positron energy ($E_e$),
photon energy ($E_\gamma$), relative time between the positron and
photon ($\tegamma$) and opening angles between the two particles
($\theta_{e \gamma}$ and $\phi_{e \gamma}$).

After the opening of the   blinding-box  the number of
$\mu \to e \gamma$ events is determined by means of a maximum likelihood fit
in the analysis  region defined as $48\,{\rm
MeV}<E_{\gamma}<58\,{\rm MeV}$, $50\,{\rm MeV}<E_{e}<56\,{\rm
MeV}$, $|\tegamma|<0.7\,{\rm ns}$, $|\theta_{e \gamma}| < 50\,{\rm
mrad}$ and $|\phi_{e \gamma}| <50\,{\rm mrad}$.

An extended likelihood function $   {\cal L}(\nsig, \nrd, \nbg)$ is constructed as
\begin{equation}
\label{eq:likelihood}
 \frac{N^{\nobs}\exp^{-N}}{\nobs!}\prod_{i = 1}^{\nobs}\left[\frac{\nsig}{N}S+\frac{\nrd}{N}R+\frac{\nbg}{N}B\right] \nonumber
\end{equation}
where $\nsig$, $\nrd$ and $\nbg$ are the number of $\meg$, RMD and
accidental background (BG) events, respectively, while $S$, $R$ and
$B$ are their respective probability density functions (PDFs).
$\nobs  = 370 $ is defined as the total number of events observed
in the analysis window and  $N = \nsig + \nrd + \nbg$. The signal
PDF $S$ is the product of the statistically independent PDFs for the
five observables ($\egamma$, $\epositron$, $\tegamma$, $\dtheta$ and
$\dphi$), each defined by their corresponding detector response
function with the measured resolutions as reported in Tab. 1.

\begin{table}[hb]
\label{table:1}
\caption{ Resolution (Gaussian $\sigma$) and efficiencies. }
\newcommand{\m}{\hphantom{$-$}}
\newcommand{\cc}[1]{\multicolumn{1}{c}{#1}}
\begin{tabular}{@{}ll}
\hline
\hline
$\frac{ \sigma _{E_{e^+}} }{ E_{e^+} }$        &  \m  0.74 \% (core fraction  83 \%)  \\
$e^+$ angle        &  \m 7.1 mrad ($\phi$ core), 11.2 ($\theta$)  \\
$e^+$ vertex  position    &  \m 3.3-3.4 mm  \\
$\frac{ \sigma _{E_{\gamma}} }{ E_{\gamma} }$     ($w$ $>$ 2$cm$)     &  \m  2.1\% \\
$\gamma$ position at LXe    &  \m 5-6 mm   \\
$\gamma$-$e^+$ timing      &  \m  142 ps \\
\hline
$\gamma$     eff.   ($\epsilon_{\gamma}$)    &  \m 58\% \\
$e^+$     eff.    &  \m 40\% \\
\hline
\hline
\end{tabular}\\[2pt]
\end{table}

\begin{figure}[h]
\includegraphics[scale=0.4]{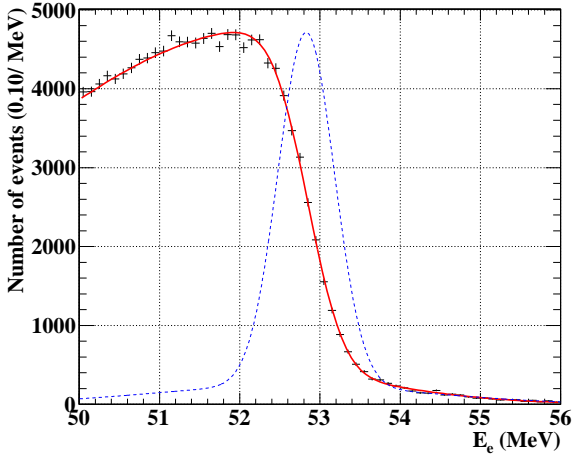}
\caption{ Positron energy spectrum with  a fit superimposed. The signal resolution function extracted with this method is shown 
(dashed).}
\label{fig:Michel}
\end{figure}

 The resolutions of the positron track reconstruction
are estimated by exploiting tracks with two full turns
in the DCH. Each turn is treated as an independent track and
 the resolutions are extracted from the 
difference between the two reconstructed tracks at the
point of closest approach to the beam axis. A fit   to the  kinematic edge of the measured Michel positron
energy spectrum  gives a cross-check for resolution and set the  absolute scale of $E_{e^+}$ (Fig.\ref{fig:Michel}).

\begin{figure}[h]
\includegraphics[scale=0.4]{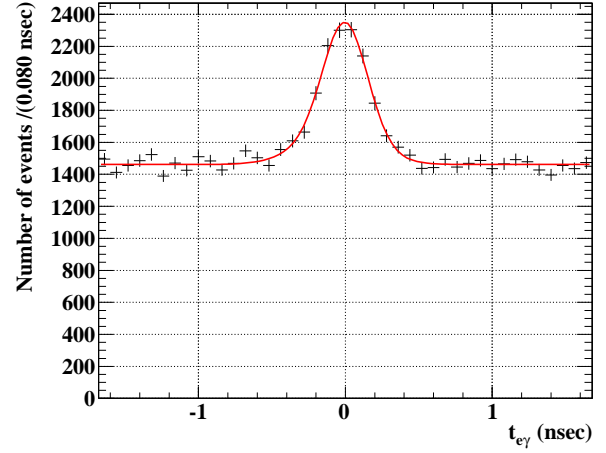}
\caption{  $\tegamma$ distribution on events with $E_{\gamma} < 48 $$MeV$. A superimposed Gaussian fit to the 
RMD peak is shown on top of the flat accidental background. }
\label{fig:tegamma}
\end{figure}

The relative time resolution at the signal
energy is estimated   from the spread of the
RMD peak observed in the region with $E_{\gamma} < 48$ $MeV$
 as shown in Fig.\ref{fig:tegamma}.

 The
RMD PDF $R$ is the product of the PDF for $\tegamma$, which is the
same as that for the signal and the PDF for the other correlated
observables ($\egamma$, $\epositron$, $\dtheta$ and $\dphi$). The
latter is formed by folding the theoretical RMD spectrum
\cite{kuno-okada} with the detector response functions. 
The BG PDF
$B$ is the product of the background spectra for the five
observables, which are precisely measured in the data sample in the
side-bands outside the blinding-box.

The number of Michel positrons counted simultaneously with the signal with the same analysis
cuts is  $\nenunu$ =18096 acquired with a dedicated trigger with a pre-scale factor $P_{e \nu \nubar}$ = 1.2 $\times$ $10^7$. 
The fraction of Michel spectrum detected in the  geometrical acceptance of the spectrometer is calculated to be $f_{e \nu \nubar}$  =0.114. 
Those numbers are used for normalization of the branching fraction of $\mu \to e \gamma$ decay
with a small corrections of momentum dependence of efficiencies:
\begin{equation}
\label{eq:norm}
\nonumber
 \frac{BF(\mu \to e \gamma)}{BF(\mu \to e \nu \nubar)} = \frac{\nsig} {\nenunu}  \times \frac{f_{e \nu \nubar}} {P_{e \nu \nubar}} \times R_{trig} \times R_e \times \epsilon_{\gamma} \\ \nonumber
 \end{equation}
 where $R_{trig}$ and $R_e$ are efficiency ratios between signal and Michel positron for trigger and positron
reconstruction respectively and are both close to 1.  $ \epsilon_{\gamma}$  is the  $\gamma$-ray detection efficiency conditional to the detection of a 
corresponding signal positron. In this data samples   $ \frac{BF(\mu \to e \gamma)}{BF(\mu \to e \nu \nubar)}  = \nsig \times (1.01 \pm 0.08) 10^{-12}$.

 We define the BF sensitivity of this search as  the mean value  of the 90\% C.L.  upper-limit distribution evaluated  over an
 ensemble of toy-MC experiments with null signal.  For this analysis it results to be  6.1 $10^{-12}$.
 This is consistent with   upper limits   obtained on sidebands samples defined by shifting the $\tegamma$ values.   The likelihood analysis is performed  on several comparable analysis 
 windows and the measured   upper limits are  in the range  (4 - 6) $10^{-12}$.

\begin{figure}[h]
\includegraphics[scale=0.4]{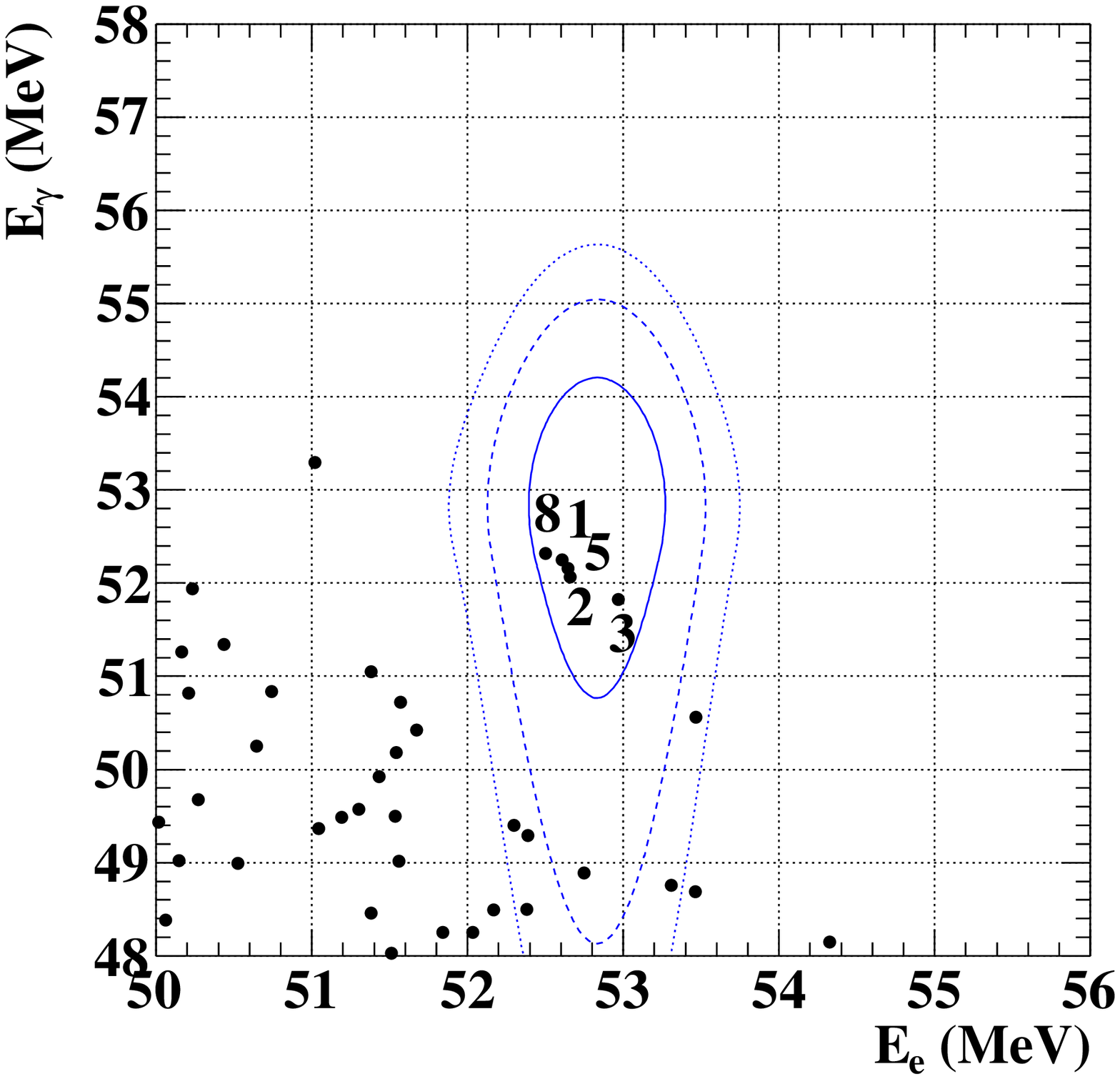}
\includegraphics[scale=0.4]{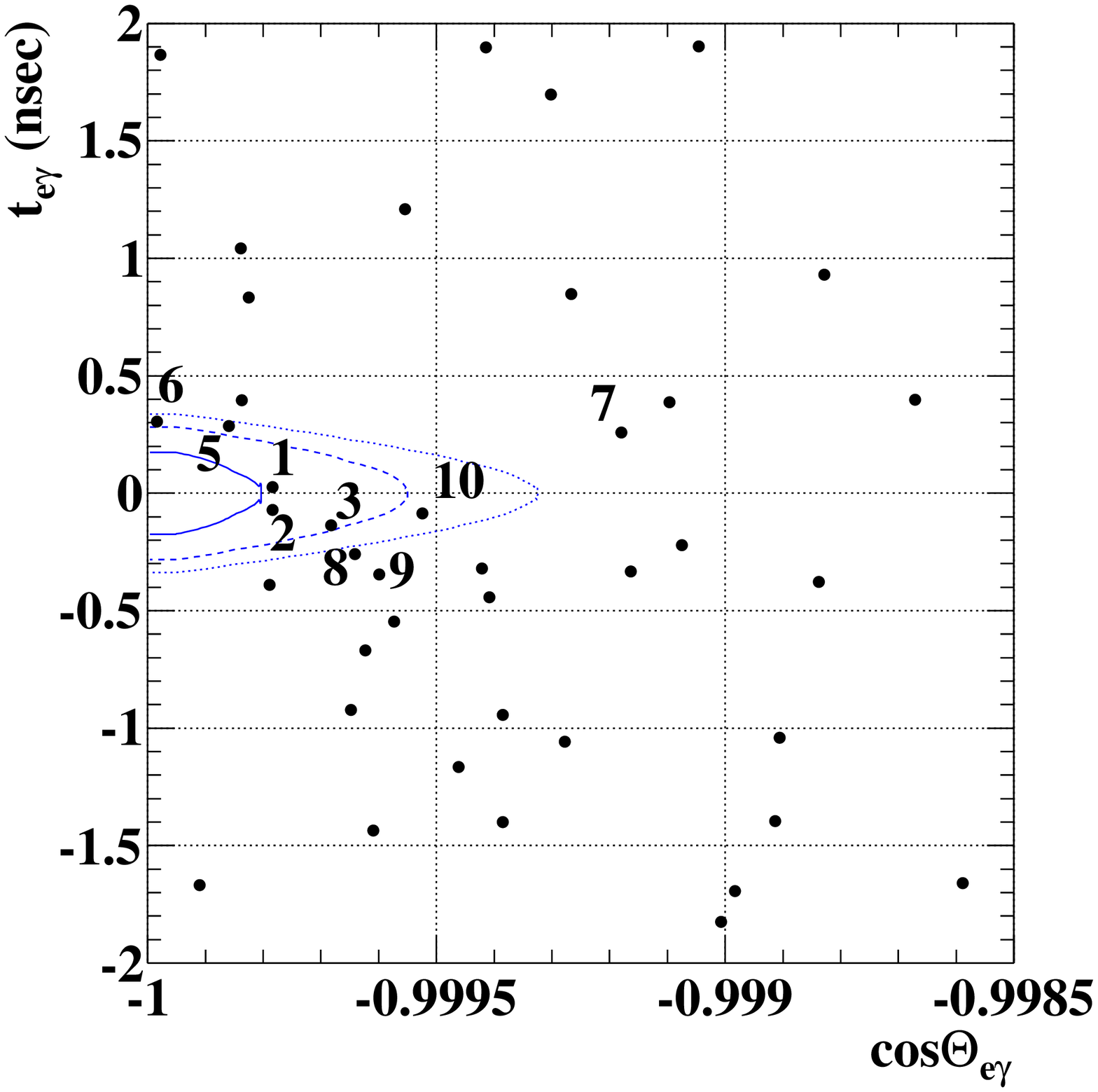}
\caption{ 
Event distribution (top) in  $E_e$-$E_{\gamma}$ plane after cuts on cos $\Theta$$_{e\gamma}$ and $\tegamma$ with 90\% efficiency 
are applied and (down) in cos $\Theta$$_{e\gamma}$-$\tegamma$  plane after cuts on $E_{e}$ and $E_{\gamma}$ with 90\% efficiency  are applied,
where $\Theta$$_{e\gamma}$  is  the opening angle between the two particle directions. The contours of signal PDFs at 1-, 1.64-
and 2-$\sigma$ are shown. Same events in the two plots are numbered correspondingly by decreasing ranking by
the relative signal likelihood (S/(R+B)).
}
\label{fig:eventranked}
\end{figure}

 Fig.\ref{fig:eventranked} shows the event
distributions inside the analysis region. The events located  close
to the signal region were carefully checked and no strange behaviour was found. The maximum likelihood fit gives  3.0 as  $\nsig$  best  value.
 The corresponding $\nrd$ is $ 35 ^{+24}_{-22}$, consistent with the expected number of  $32 \pm 2$ from the $E_{\gamma}$ sideband.  
 The confidence region is constructed by means of toy-MC simulation with taking into account possible
systematic effects. The PDFs and normalization factor are fluctuated for each toy-MC experiment
in accordance with their uncertainty values. The point of $\nsig$=0 is included in the 90\% confidence
interval, and an upper limit is calculated to be $\nsig$ $<$ 14.5. This yields an upper limit on the BF
\begin{equation}
\label{eq:result}
   \frac{BF(\mu \to e \gamma)}{BF(\mu \to e \nu \nubar)}  < 1.5 \times 10^{-11}  at  90 \% C.L. 
\end{equation}
Three independent analyses with different statistical approaches were performed to check the analysis, and gave consistent results.
 
\section{CONCLUSIONS and PERSPECTIVES}

A search for the LFV decay $\megsign$ was
performed with a branching ratio sensitivity of $6.1\times10^{-12}$,
using data taken during 2009. With this sensitivity  a
blind likelihood analysis yields an upper limit on the branching
ratio of BR$({\mu^+ \to e^+ \gamma})  <  1.5 \times 10^{-11}$ (90\%
C.L.).  MEG resumed data-taking in 2010 and is accumulating data with detector stable condition.
 It will run until the end of 2012 to reach the sensitivity of few $10^{-13}$.

\end{document}